\documentclass[aps,pra,twocolumn,floats]{revtex4-2} 
\usepackage{amsmath}
\usepackage{graphicx}
\usepackage{dcolumn}
\usepackage{bm}
\usepackage{amssymb}
\usepackage{latexsym}
\usepackage{color}
\usepackage{subfigure}

\definecolor{purple}{rgb}{0.5, 0.0, 0.5}

\newcommand{\veps}{\varepsilon}

\usepackage{ulem}

\def\1S0{${}^{1}$S$_{0}$}
\def\3P1{${}^{3}$P$_{1}$}

\begin{document}

\title{Exotic Kondo effect in two one dimensional spin 1/2 chains coupled to two localized spin 1/2 magnets}

\author{Igor Kuzmenko$^{1,2}$, Tetyana Kuzmenko$^1$, Y. B. Band$^{1,2,3}$ and Yshai Avishai$^{1,4}$}

\affiliation{
  $^1$Department of Physics,
  Ben-Gurion University of the Negev,
  Beer-Sheva 84105, Israel
  \\
  $^2$Department of Chemistry,
  Ben-Gurion University of the Negev,
  Beer-Sheva 84105, Israel
  \\
  $^3$The Ilse Katz Center for Nano-Science,
  Ben-Gurion University of the Negev,
  Beer-Sheva 84105, Israel
  \\
  $^4$Yukawa Institute of Theoretical Physics, Kyoto, Japan
  }

\begin{abstract}
We study an exotic Kondo effect in a system consisting of two one-dimensional XX Heisenberg ferromagnetic spin $1/2$ chains (denoted by $\alpha = u, d$ for up and down chains) coupled to a quantum dot consisting of two localized spin $1/2$ magnets. Using the Jordan-Wigner transformation on the Heisenberg Hamiltonian of the two chains, this system can be expressed in terms of non-interacting spinless fermionic quasiparticles.  As a result, the Hamiltonian of the whole system is expressed as an Anderson model for spin 1/2 fermions interacting with a spin-1/2 impurity.  Thus, we study the scattering of fermionic quasiparticles (propagating along spin chains) by a pair of localized magnetic impurities.  At low temperature, the localized spin $1/2$ magnets are shielded by the chain `spins' via the Kondo effect. We calculate the Kondo temperature $T_K$ and derive the temperature dependence of the entropy, the specific heat, the specific heat and the `magnetic susceptibility' of the dot for $T \gg T_K$.  Our results can be generalized to the case of anti-ferromagnetic XX chains.
\end{abstract}

\maketitle

\section{Introduction}
  \label{sec:intro}

The physics of low-dimensional magnetic systems, and in particular an array of one-dimensional magnets, is an active area of solid state physics.  Such materials offer a unique opportunity to study ground and excited states of quantum magnetic arrays, possible new phases of matter, and the interplay between quantum and thermal fluctuations. For more details see, e.g., Ref.~\cite{Mikeska-1D-Heisenberg-2004}. A quasi-one dimensional $S = 1/2$ antiferromagnet is observed in CuCl$_2 \cdot 2$NC$_5$H$_5$ (copper pyridine chloride) \cite{Heilmann-PRB-1978}, CuGeO$_3$ \cite{Hase-PRL-1993}, Ba$_3$Cu$_3$In$_4$O$_{12}$ and Ba$_3$Cu$_3$Sc$_4$O$_{12}$ \cite{Dutton-JPhysCondMat-2012}.  The quantum spin dynamics in a ferromagnetic Ising chain material CoNb$_2$O$_6$ is studied in Ref.~\cite{Amelin-Ising-2020}.  As shown in Ref.~\cite{Bloch-cold-atoms-RevModPhys-2008}, it is possible to realize a tunable isotropic or anisotropic Heisenberg model with ultracold atoms in optical lattices.  The Heisenberg spin $1/2$ chain with a concentration $x$ of impurity spins $S$ subject to an external magnetic field is theoretically studied in Ref.~\cite{Schlottmann-spin-chain-impurities-PRB-1994} using the Bethe thermodynamic ansatz. In the absence of an external magnetic field, the ground state is a singlet for finite $x$, but non-Fermi liquid-like for $x \to 0$.  Reference~\cite{Kattel-spin-Kondo-PRB-2024} studies the boundary effects that arise when spin 1/2 impurities interact with the edges of an anti-ferromagnetic spin 1/2 Heisenberg chain through spin-exchange interactions. It shows that in the case of anti-ferromagnetic interaction, when the impurity coupling strength is much weaker than in the bulk, the impurity is shielded in the ground state by the Kondo effect.

The original Kondo effect describes the scattering of conduction electrons by magnetic impurities in dilute magnetic alloys~\cite{Kondo-ProgrTheorPhys-1964, Hewson}. The anti-ferromagnetic exchange interaction between the magnetic impurities and the conduction electrons leads to a minimum of the electrical resistivity with temperature~\cite{Hewson}. The low-energy spin waves are studied in Ref.~\cite{Bao-magnon-Kondo-2023} in a metallic van der Waals ferromagnet Fe$_{3-x}$GeTe$_2$ (Curie temperature $T_C = 160$~K), where the Kondo lattice behavior appears in the ferromagnetic phase below a characteristic temperature $T^* \approx 90$~K. It is shown that the magnon damping constant exhibits a minimum around $T^*$, which is analogous to the resistivity minimum due to the single-impurity Kondo effect~\cite{Bao-magnon-Kondo-2023}.

\begin{figure}
\centering
  \includegraphics[width = 0.65 \linewidth,angle=0] {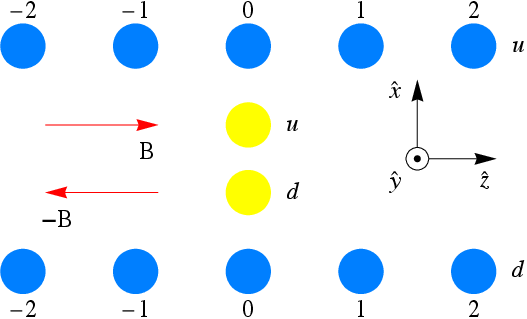}
\caption{\footnotesize  
Ferromagnetic Heisenberg XX chains  (azure) coupled to local magnetic moments (yellow).  The system is placed in the $x$-$z$ plane.  The chain and the dot, which are located in the half-plane $x > 0$ ($x < 0$), are labeled by the index $\alpha = u$ ($\alpha = d$).  Furthermore, the sites of the chain are labelled by an integer $n$.  An external magnetic field $\mathbf{B} (x) = B \, \mathrm{sign} (x) \, \hat{\mathbf{z}}$ is applied.  This field is used in Sec. ~\ref{sec:susceptibility} titled ``Magnetic Susceptibility''.
\label{Fig:chains-dot}}
\end{figure}

Here we present a novel model for the Kondo effect, which describes scattering of Jordan-Wigner fermionic quasiparticles (propagating in 1D ferromagnetic chains), on a quantum dot composed of two localized spin 1/2 magnetic impurities. This kind of Kondo effect is visualized in Fig.~\ref{Fig:chains-dot}, that contains the following elements:
\begin{itemize}
\item
Two spin 1/2 XX Heisenberg chains labeled $\alpha = u, d$, and the sites of the chain are labelled by an integer $n$;

\item
Two localized quantum magnets with Ising coupling labeled $\alpha = u, d$;

\item
XX interaction between the localized magnets and the spin chains.

\end{itemize}
First we shall calculate thermodynamic quantities without an external magnetic field.   Later on, in Sec.~\ref{sec:susceptibility}, an external magnetic field is applied directed along $\hat{\bf z}$ in the `upper' domain ($u$)  and along $-\hat{\bf z}$ in the `lower' domain ($d$), in order to calculate  ``Magnetic Susceptibility''.

The Jordan-Wigner transformation (JWT)~\cite{Mikeska-1D-Heisenberg-2004} enables representation of the system Hamiltonian as an Anderson model Hamiltonian~\cite{Anderson-PRB-1961} for spin 1/2 fermions interacting with a single spin 1/2 impurity.  At low temperature, the impurity spin is screened by the chain spins via the Kondo effect. 

The paper is organized as follows.  Section~\ref{sec:system} introduces the theoretical model of two Heisenberg spin 1/2 chains interacting with two localized magnetic impurities.  In Sec.~\ref{sec:Jordan-Wigner},  the JWT is applied, resulting in an expression for the spin Hamiltonian as an Anderson model Hamiltonian.  The Schrieffer-Wolff transformation is introduced in Sec.~\ref{sec:Schrieffer-Wolff}, which determines an effective Kondo Hamiltonian.  In Sec.~\ref{sec:Kondo-temperature}  we employ the `poor man's scaling' technique that enables elucidation of the Kondo effect and compute the Kondo temperature $T_K$, which is the scaling invariant of the RG equation. In section \ref{sec:entropy-specific-heat} we compute the temperature dependence of the entropy and the specific heat for $T \gg T_K$.  In Sec.~\ref{sec:susceptibility} we derive the zero-field `magnetic susceptibility' of the dot. The results are summarized in Sec.~\ref{sec:conclusion}.

\section{Description of the system} \label{sec:system}

We consider two ferromagnetic Heisenberg XX spin chains coupled to two localized magnetic moments, which we refer to as {\it dots} or {\it impurities}, as illustrated in Fig.~\ref{Fig:chains-dot}.  The system is positioned in the $x$-$z$ plane.  The chain and the localized magnet situated in the $x > 0$ half-plane are designated by the index $\alpha = u$, and the chain and the localized magnet placed in the $x < 0$ half-plane are labeled by $\alpha = d$. Furthermore, the sites of the chains are enumerated by integer $n$, with $|n| \leq N$, where the number of sites of each chain is $2 N + 1$. The Hamiltonian of the system is given by the sum of three terms:
\begin{equation}   \label{eq:H=H0+HD+Hint}
  H = H_0 + H_D + H_{\mathrm{int}} .
\end{equation}
The first term, $H_0$, is the Hamiltonian of two ferromagnetic chains,
\begin{eqnarray}
  H_0 &=&
  - v
  \sum_{\alpha}
  \sum_{n = - N}^{N}
  \Big[
    s_{\alpha, x} (n) \,
    s_{\alpha, x} (n + 1)
    \nonumber \\ && +
    s_{\alpha, y} (n) \,
    s_{\alpha, y} (n + 1)
  \Big] .
  \label{eq:H0}
\end{eqnarray}
For periodic boundary conditions $n$ goes from $-N$ to $N$ where $\mathbf{s}_{\alpha} (n) = ( s_{\alpha, x} (n), s_{\alpha, y} (n), s_{\alpha, z} (n))$ represents a spin-$1/2$ operator of a site $n$ in a chain $\alpha$, and $v > 0$ is a ferromagnetic coupling strength. The case of anti-ferromagnetic interaction, where $v < 0$, is discussed after Eq.~(\ref{eq:energy-fermion}). Moreover, we assume the periodic boundary conditions ${\bf s} (n + 2 N + 1) = {\bf s} (n)$, hence the number of sites is $2 N + 1$.

Note that inclusion of a $z$-component in the Hamiltonian (\ref{eq:H0}) would result in a Hamiltonian that is called the XXZ Heisenberg anisotropic Hamiltonian, and the physics would be very different from that obtained below.
The XXZ  model Hamiltonian is derived from the XX Hamiltonian in Eq.~(\ref{eq:H0}) by the addition of a term, $\Delta \, s_{\alpha, z} (n) s_{\alpha, z} (n + 1)$.  For $\Delta > 1$, the ground state is ferromagnetic, and for $\Delta < - 1$, the ground state is an anti-ferromagnetic Ising or N\'eel phase~\cite{Mikeska-1D-Heisenberg-2004}. In both of these cases, the excitation spectrum has a gap of $J (|\Delta| - 1)$~\cite{Mikeska-1D-Heisenberg-2004}. At $|\Delta| = 1$, the chains have continuous rotational symmetry and the spectrum becomes gapless.  For $|\Delta| < 1$, the ground state of the XXZ chains is the XY phase characterized by uniaxial symmetry of the easy-plane type and a gapless excitation continuum~\cite{Mikeska-1D-Heisenberg-2004}. The Jordan-Wigner transformation (discussed in Sec.~\ref{sec:Jordan-Wigner} below) expresses the XXZ Hamiltonian in terms of interacting 1D fermions.  We discuss here mainly the simplest case, $\Delta = 0$ (the XX model), where the fermions are non-interacting.  The case of interacting fermions will be considered in a future publication.

The second term on the right hand side of Eq.~(\ref{eq:H=H0+HD+Hint}), $H_D$, describes the two localized quantum magnets (`dots'),
\begin{eqnarray}
  H_D &=& w \, \sigma_{u,z} \sigma_{d, z} ,
  \label{eq:HD}
\end{eqnarray} 
where $\boldsymbol\sigma_{\alpha} = ( \sigma_{\alpha, x}, \sigma_{\alpha, y}, \sigma_{\alpha, z} )$ represents a spin $1/2$ operator of the localized magnet, and $w > 0$ is an {\it anti-ferromagnetic} coupling strength.

The last term on the RHS of Eq.~(\ref{eq:H=H0+HD+Hint}) describes the interaction between the chains and the localized magnets,
\begin{equation}   \label{eq:Hint}
  H_{\mathrm{int}} =
  - \frac{t}{2}
  \sum_{\alpha}
  \big[
    s_{\alpha}^{+} (0) \, \sigma_{\alpha}^{-} +
    s_{\alpha}^{-} (0) \, \sigma_{\alpha}^{+}
  \big] ,
\end{equation}
where $s_{\alpha}^{\pm} (n) = s_{\alpha, x} (n) \pm i s_{\alpha, y} (n)$,
$\sigma_{\alpha}^{\pm} = \sigma_{\alpha, x} \pm i \sigma_{\alpha, y}$, and
the coupling constant $t$ is real (Note that $t$ can be either positive or negative.)

\section{Jordan-Wigner transformations}
  \label{sec:Jordan-Wigner}

The Hamiltonian $H_0$ in Eq.~(\ref{eq:H0}) can be rewritten in terms of fermionic creation and annihilation operators using the JWT~\cite{Mikeska-1D-Heisenberg-2004}:
\begin{eqnarray}
  s_{\alpha}^{+} (n) &=&
  c_{\alpha}^{\dag} (n) \,
  U_{\alpha} (n) ,
  \label{eq:Jordan-Wigner-chains}
  \\
  \sigma_{\alpha}^{+} &=&
  \gamma_{\alpha}^{\dag} \,
  \mathcal{U}_{\alpha} ,
  \label{eq:Jordan-Wigner-dot}
\end{eqnarray}
where $c_{\alpha} (n)$ and $\gamma_{\alpha}$ are fermionic annihilation operators.  Here the unitary operators $U_{\alpha} (n)$ and $\mathcal{U}_{\alpha}$ are defined as
\begin{eqnarray}
  U_d (n) &=&
  \exp
  \bigg(
    i \pi \sum_{n' = - N}^{n - 1}
    c_{d}^{\dag} (n') \, c_{d} (n')
  \bigg) ,
  \label{eq:U_L-chain}
  \\
  U_u (n) &=&
  U_d (N+1) \,
  \exp
  \bigg(
    i \pi \sum_{n' = - N}^{n - 1}
    c_{u}^{\dag} (n') \, c_{u} (n')
  \bigg) ,
  \label{eq:U_R-chain}
  \\
  \mathcal{U}_d &=&
  U_u (N+1) ,
  \label{eq:U_L-dot}
  \\
  \mathcal{U}_u &=&
  \mathcal{U}_d \,
  \exp
  \big(
    i \pi \gamma_{d}^{\dag} \, \gamma_d
  \big) ,
  \label{eq:U_R-dot}
\end{eqnarray}
and the spin-ladder operators $s_{\alpha}^{\pm} (n)$ and $\sigma_{\alpha}^{\pm}$ are given by
\begin{eqnarray}
  s_{\alpha}^{\pm} (n) &=&
  s_{\alpha, x} (n) \pm i s_{\alpha, y} (n) ,
  \label{eq:spin-ladder-chain}
  \\
  \sigma_{\alpha}^{\pm} &=&
  \sigma_{\alpha, x} \pm i \sigma_{\alpha,, y} .
  \label{eq:spin-ladder-dot}
\end{eqnarray}
These relations are already defined following Eq.(4).  Employing Eq.~(\ref{eq:Jordan-Wigner-chains}), Eq.~(\ref{eq:H0}) can be written as,
\begin{eqnarray}
  H_0 &=&
  - \frac{v}{2}
  \sum_{\alpha}
  \sum_{n = -N}^{N}
  \big[
    c_{\alpha}^{\dag} (n) \,
    c_{\alpha} (n + 1) +
    \mathrm{h.c.}
  \big] .
  \label{eq:h0-fermionic}
\end{eqnarray}
This is a 1D tight-binding model Hamiltonian for fermions.  In other words, the Jordan-Wigner transformations generate spinless fermions, while the chain index $\alpha$ in Eq.~(\ref{eq:h0-fermionic}) plays the role of pseudo-spin.

The Hamiltonian $H_0$ can be diagonalized using the Fourier transformation
\begin{equation}   \label{eq:Fourier-c}
  c_{\alpha} (n) =
  \frac{1}{\sqrt{2 N + 1}}
  \sum_{k}
  c_{\alpha, k} \,
  e^{- i k n} ,
\end{equation}
where $c_{\alpha, k}$ is an annihilation operator of fermion with spin $\alpha$ and wave number $k$ belonging to the first BZ of the reciprocal lattice, $- \pi < k \leq \pi$.  The transformed Hamiltonian is
\begin{equation}   \label{eq:H0-diagonal}
  H_0 =
  \sum_{k, \alpha}
  \epsilon_k \,
  c_{k \alpha}^{\dag}
  c_{k \alpha} ,
\end{equation}
where the fermion energy is
\begin{equation}   \label{eq:energy-fermion}
  \epsilon_k = - v \, \cos k .
\end{equation}

The energy band  $\epsilon_k$ is an even function of $k$, i.e., $\epsilon_{-k} = \epsilon_k$.  If $v > 0$, then the minimum value of $\epsilon_k$ is attained at $k = 0$, with $\epsilon_0 = - v$.  The maximum value is attained at $k = \pm \pi$, with the value of the energy at this point given by  $\epsilon_{\pm \pi} = v$, and the bandwidth is $2 v$.  Furthermore, $\epsilon_{\pm \pi/2} = 0$ is the Fermi energy.  In the ground state of the chains, all the energy levels with $\epsilon_k < 0$ (i.e., $|k| < \pi/2$) are occupied, while the levels with $\epsilon_k > 0$ (and $\pi/2 < |k| < \pi$) are empty.  For anti-ferromagnetic coupling (i.e., if, $v < 0$), the energy band $\epsilon_k$ has minima for $k = \pm \pi$, maximum for $k = 0$.  The occupied energy levels have $\pi/2 < |k| < \pi$, and  the bandwidth is $2 |v|$.

The density of states (DOS)  of the chains, $\rho (\epsilon)$, is given by
\begin{equation}   \label{eq:DOS-def}
  \rho (\epsilon) =
  \frac{1}{2 N + 1}
  \sum_{k}
  \delta (\epsilon(k) - \epsilon) ,
\end{equation}
where $\delta (\bullet)$ is the Dirac delta function.  In the limit $N \to \infty$, the sum can be approximated by an integral as
$$
  \frac{1}{2 N + 1} \sum_{k} \cdots
  \approx \frac{1}{2 \pi} \int\limits_{-\pi}^{\pi} \cdots \, d k.
$$
The DOS in this limit takes the form
\begin{equation}   \label{eq:DOS-res}
  \rho (\epsilon) =
  \frac{\Theta (v - |\epsilon|)}{\pi \, \sqrt{v^2 - \epsilon^2}} ,
\end{equation}
when $v > 0$, and for $v < 0$, $\frac{\Theta (|v| - |\epsilon|)}{\pi \, \sqrt{v^2 - \epsilon^2}}$, where $\Theta(\bullet)$ is the Heaviside step function.  Note that if $|\epsilon| \to |v|$, the DOS diverges as $\rho (\epsilon) \sim (|v| - |\epsilon|)^{-1/2}$, and if $\epsilon = \epsilon_F$, $\rho (\epsilon_F) = 1/(\pi |v|)$.

Using Eq.~(\ref{eq:Jordan-Wigner-dot}), the dot Hamiltonian $H_D$ in Eq.~(\ref{eq:HD}) can be written as,
\begin{equation}   \label{eq:hD-fermionic}
  H_D =
  - \frac{w}{2} \sum_{\alpha} n_{\alpha} + w \, n_u \, n_d ,
\end{equation}
where $n_{\alpha} = \gamma_{\alpha}^{\dag} \gamma_{\alpha}$ is the number operator of fermions with pseudo-spin $\alpha$.  According to the Pauli principle, the dot can be either unoccupied (where $n_u + n_d = 0$), or occupied by one fermion (where $n_u + n_d = 1$), or by two fermions (where $n_u + n_d = 2$).  The wave function $| 0 \rangle$ of the unoccupied quantum dot satisfies the equality $\gamma_{\alpha} | 0 \rangle = 0$.  The wave functions of the singly occupied dot are $| \alpha \rangle = \gamma_{\alpha}^{\dag} | 0 \rangle$, and the wave function of the doubly occupied dot is $| s \rangle = \gamma_{u}^{\dag} \gamma_{d}^{\dag} | 0 \rangle$ (where the index $s$ denotes the `singlet' state, see Sec. ~\ref{sec:Schrieffer-Wolff} below).  The ground state is a twofold degenerate singlet with energy $\veps_u = \veps_d = - w/2$, and the excited state is a twofold degenerate with energy $\veps_0 = \veps_s = 0$.

Finally, using Eqs.~(\ref{eq:Jordan-Wigner-chains}) and (\ref{eq:Jordan-Wigner-dot})
on Eq.~(\ref{eq:Hint}), we can express $H_{\mathrm{int}}$ as
\begin{equation}   \label{eq:Hint-fermion}
  H_{\mathrm{int}} =
  - \frac{t}{2}
  \sum_{\alpha}
  \big[
    c_{\alpha}^{\dag} (0) \, \gamma_{\alpha} +
    \gamma_{\alpha}^{\dag} \, c_{\alpha} (0)
  \big] .
\end{equation}
Applying the Fourier transform in Eq.~(\ref{eq:Fourier-c}), we can write $H_{\mathrm{int}}$ as
\begin{equation}   \label{eq:Hint-fermion-Fourier}
  H_{\mathrm{int}} =
  - \frac{t}{2 \sqrt{2 N + 1}}
  \sum_{k, \alpha}
  \big[
    c_{k, \alpha}^{\dag} \, \gamma_{\alpha} +
    \gamma_{\alpha}^{\dag} \, c_{k, \alpha}
  \big] .
\end{equation}

\section{Schrieffer-Wolff transformation}
  \label{sec:Schrieffer-Wolff}

The Hamiltonian $H$ given in Eqs.~(\ref{eq:H=H0+HD+Hint}), (\ref{eq:H0-diagonal}), (\ref{eq:hD-fermionic}) and (\ref{eq:Hint-fermion-Fourier}) is, in fact, an Anderson impurity model Hamiltonian,  which is similar to a model used to describe magnetic impurities embedded in metals \cite{Anderson-PRB-1961}.  When the temperature $k_B T$ is much smaller than the quantum dot excitation energy $w$, we can apply the Schrieffer-Wolff transformation to project out the high energy excitations $| 0 \rangle$ and $| s \rangle$ of the Anderson impurity model Hamiltonian and obtain an effective low energy model Hamiltonian \cite{Bravyi-Schrieffer-Wolf-AnnPhys-2010}:
\begin{equation}   \label{eq:H=H0+HK}
  H = H_0 + H_K .
\end{equation}
Here $H_0$  is given in Eq.~(\ref{eq:H0-diagonal}), and $H_K$ has the form of an $s$-$d$ exchange interaction Hamiltonian used in 
the Kondo model \cite{Kondo-ProgrTheorPhys-1964},
\begin{equation}   \label{eq:HK}
  H_K = J \, \mathbf{S} \cdot \boldsymbol\Sigma ,
\end{equation}
where $\mathbf{S}$ is the local pseudo-spin density operator of the chains, $\boldsymbol\Sigma$ is a pseudo-spin operator of the dot, and $J$ is the exchange coupling:
\begin{eqnarray}
  \mathbf{S} &=&
  \frac{1}{2 (2 N + 1)}
  \sum_{k, k', \alpha, \alpha'}
  c_{k, \alpha}^{\dag}
  \boldsymbol\tau_{\alpha, \alpha'}
  c_{k', \alpha'} ,
  \label{eq:pseudospin-chain-def}
  \\
  \boldsymbol\Sigma &=&
  \frac{1}{2}
  \sum_{\alpha, \alpha'}
  \gamma_{\alpha}^{\dag}
  \boldsymbol\tau_{\alpha, \alpha'}
  \gamma_{\alpha'} ,
  \label{eq:pseudospin-dot-def}
  \\
  J &=& \frac{4 t^2}{w} .
  \label{eq:J-def}
\end{eqnarray}
Here $\boldsymbol\tau = (\tau_x, \tau_y, \tau_z)$, which appears in both $\mathbf{S}$ and $\boldsymbol\Sigma$, is a vector of Pauli matrices.  Note that the operators 
$\mathbf{S}$ and $\boldsymbol\Sigma$ are not `real' spin operators, that is, they do not represent an intrinsic angular momentum carried by particles. In this context, fermion quasiparticles with $\alpha = d$ are designated as `pseudo-spin-down' quasiparticles, and $\alpha = u$ as `pseudo-spin-up' quasiparticles. The operators $\mathbf{S}$ and $\boldsymbol\Sigma$ permit `pseudo-spin-flipping' transitions, and satisfy the commutation relations of spin operators, therefore they can be considered as pseudo-spin 1/2 operators.

\section{Scaling equation and Kondo temperature}  \label{sec:Kondo-temperature}

In order to derive scaling equations we define a dimensionless coupling $j = J \rho (\epsilon_F)$.  Within the standard poor man's scaling technique, $j(D)$ is renormalized as the original half-bandwidth $D_0 = v$ and is reduced to $D < D_0$ by integrating out high energy excitations~\cite{Hewson}.  The scaling equation for $j (D)$ supported by the initial condition at $D = D_0$ reads
\begin{eqnarray}
  \frac{d j(D)}{d \ln D} &=& - j^2 (D) ,
  \label{eq:scaling}
  \\
  j (D_0) &=&\frac{4 t^2}{\pi v w} .
  \label{eq:j0ini}
\end{eqnarray}
Equation (\ref{eq:scaling}) has the solution
\begin{equation}   \label{eq:j-vs-D}
  j (D) = \frac{1}{\ln \big( D / T_K \big)} ,
\end{equation}
where the Kondo temperature $T_K$ (the scaling invariant of the RG equation) is given by
\begin{equation}   \label{eq:TK}
  T_K = \frac{v}{k_B} \, \exp \Big( - \frac{\pi v w}{4 t^2} \Big) .
\end{equation}

\begin{figure}
\centering
  \includegraphics[width = 0.85 \linewidth,angle=0] {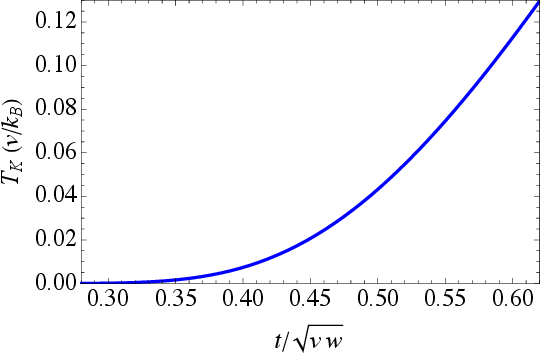}
\caption{\footnotesize  
  The Kondo temperature $T_K$ versus $t/\sqrt{v w}$, where $v$ is a half-bandwidth in Eq.~(\ref{eq:energy-fermion}),
  $w$ is an energy parameter of the dot Hamiltonian in Eq.~(\ref{eq:hD-fermionic}),  and
  $t$ is the tunneling rate in the Hamiltonians $H_{\rm int}$  in Eq.~(\ref{eq:Hint-fermion-Fourier}).
 }
\label{Fig:TK}
\end{figure}

Figure~\ref{Fig:TK} plots $T_K$ versus $t/\sqrt{v w}$.  Note that if $t < 0.34 \sqrt{v w}$, then $T_K < 0.001 \, v$ is very small. $T_K$ increases fast with $t$, and $T_K = 0.13 \, v$ for $t = 0.62 \, \sqrt{v w}$.

\section{Entropy and specific heat}  \label{sec:entropy-specific-heat}

The dependence of the entropy $\mathcal{S} (T)$ and the specific heat $\mathcal{C} (T) = T \, d \mathcal{S} (T) / d T$
of the quantum dot at $T \gg T_K$ are specified in Ref.~\cite{Hewson} as,
\begin{eqnarray}
  \mathcal{S} (T) &=&
  k_B \, \ln 2 -
  \frac{\pi^2 \, k_B}{4 \, \ln^3 (T / T_K)} ,
  \label{eq:entropy}
  \\
  \mathcal{C} (T) &=&
  \frac{3 \, \pi^2 \, k_B}{4 \, \ln^4 (T / T_K)} ,
  \label{eq:specific-heat}
\end{eqnarray}
where $k_B$ is the Boltzmann constant.  Note that the scaling technique for $\mathcal{S} (T)$ and $\mathcal{C} (T)$
diverge at $T \to T_K$.  This is due to the fact that the poor man's scaling technique in Eq.~(\ref{eq:scaling})
is derived for $T \gg T_K$, and therefore cannot be used as $T \to T_K$.

\begin{figure}
\centering
  \includegraphics[width = 0.8 \linewidth,angle=0] {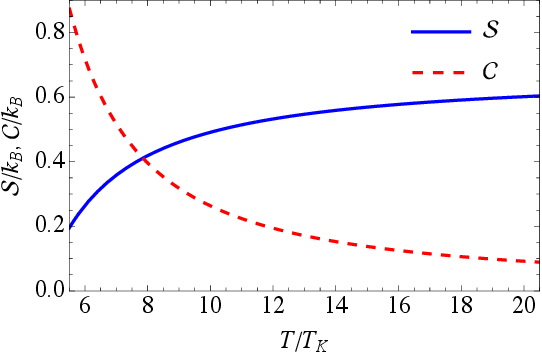}
\caption{\footnotesize  
  The entropy $\mathcal{S}(T)$ (solid blue) and
  the specific heat $\mathcal{C}(T)$ (dashed red)
  of the dot versus $T/T_K$,where the Kondo temperature $T_K$ is given in Eq.~(\ref{eq:TK})
  }
\label{Fig:SC}
\end{figure}

The entropy and the specific heat are plotted versus temperature in Fig.~\ref{Fig:SC}.  Note that as $T \to \infty$, $\mathcal{S} (T) \to \ln (2) \approx 0.69$ and $\mathcal{C} (T) \to 0$.  $\mathcal{S} (T)$ decreases with $1/T$, and $\mathcal{C} (T)$ increases with $1/T$.  When $T = 6 \, T_K$, $\mathcal{S} (T) = 0.26 \, k_B$ and $\mathcal{C} (T) = 0.72 \, k_B$.

\section{Magnetic susceptibility}  \label{sec:susceptibility}

As discussed above, the pseudo-spin operators $\mathbf{S}$ and $\boldsymbol\Sigma$ do not represent an intrinsic angular momentum carried by particles. Rather, they are operators that permit `pseudo-spin-flipping' transitions when the quasiparticles jump between the chains.  This is evident from the discussions that followed Eq.~(\ref{eq:J-def}).  Nevertheless, we will demonstrate below that the interaction of the chains and the dot with an inhomogeneous external magnetic field $\mathbf{B} (x) = B \, \mathrm{sign} (x) \, \hat{\mathbf{z}}$ shown in Fig.~\ref{Fig:chains-dot} is described by a Zeeman-type Hamiltonian that accounts for the interaction of the pseudo-spin operators $\mathbf{S}$ and $\boldsymbol\Sigma$ with a homogeneous magnetic field $B \, \hat{\bf z}$. 

Consider two spin chains coupled to two localized magnetic moments and placed in an external magnetic field $\mathbf{B} (x) = B \, \mathrm{sign} (x) \, \hat{\mathbf{z}}$, where $\mathrm{sign} (x)$ is the sign function whose value is $-1$ for $x < 0$, $1$ for $x > 0$, and $0$ for $x = 0$.  The chain and the localized magnetic moment, denoted by $\alpha = u(d)$, are placed in the half-plane with positive $x$-coordinate (negative $x$-coordinate), as shown in Fig.~\ref{Fig:chains-dot}.  The Hamiltonian of the chains is $H + H_Z$, where $H$ is given by (\ref{eq:H=H0+HK}), (\ref{eq:H0-diagonal}) and (\ref{eq:HK}), and $H_Z$ by the equation
\begin{eqnarray}
  H_Z &=&
  -g \mu_B B
  \sum_{n = -N}^{N}
  \big[
    s_{u, z} (n) -
    s_{d, z} (n)
  \big]
  \nonumber \\ &&
  -g \mu_B B \,
  \big[
    \sigma_{u, z} -
    \sigma_{d, z}
  \big] .
  \label{eq:HZ}
\end{eqnarray}
Here $g$ is the $g$-factor of the magnetic moments, and $\mu_B$ is the Bohr magneton.  Applying the JWT in Eqs.~(\ref{eq:Jordan-Wigner-chains}) and (\ref{eq:Jordan-Wigner-dot}),  $H_Z$ can be expressed as
\begin{equation}   \label{eq:HZ-fermion}
  H_Z =
  - 2 g \mu_B B
  \sum_{n = -N}^{N}
  S_Z (n)
  - 2 g \mu_B B \, \Sigma_z ,
\end{equation}
where $\Sigma_z$ is the pseudo-spin operator defined in Eq.~(\ref{eq:pseudospin-dot-def}), and the pseudo-spin operators of the chains are
\begin{equation}   \label{eq:pseudospin-chain-coord}
  \mathbf{S} (n) =
  \frac{1}{2}
  \sum_{\alpha, \alpha'}
  c_{\alpha}^{\dag} (n) \,
  \boldsymbol\tau_{\alpha, \alpha'} \,
  c_{\alpha'} (n).
\end{equation}
$H_Z$ in Eq.~(\ref{eq:HZ-fermion}) has a form of Zeeman interaction of the `magnetic moment' ${\bf M} = 2 g \mu_B {\bf S}$ with external `magnetic field' $\boldsymbol{\mathcal{B}} = B \, \hat{\mathbf{z}}$.

\begin{figure}[htb]
\centering
  \includegraphics[width = 0.8 \linewidth,angle=0] {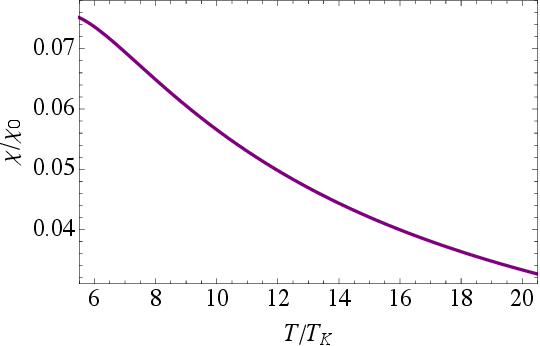}
\caption{\footnotesize  
  The magnetic susceptibility $\chi$ versus $T/T_K$, where the Kondo temperature $T_K$ is given in Eq.~(\ref{eq:TK}), and
  $\chi_0$ is given in Eq.~(\ref{eq:chi_0}). }
\label{Fig:suscept}
\end{figure}

The zero-field `susceptibility' of the dot is given by~\cite{Hewson}
\begin{equation}   \label{eq:susceptibility}
  \chi (T) =
  \frac{(2 g \mu_B)^{2}}{4 k_B T}
  \bigg(
    1 - \frac{1}{\ln (T / T_K)}
  \bigg) .
\end{equation}
At $T \gg T_K$, the logarithmic term describes screening of the dot spin due to the anti-ferromagnetic coupling $H_K$. At $T \to T_K$, the logarithmic correction to the susceptibility diverges, since the poor man's scaling technique cannot be used as $T \to T_K$.

The magnetic susceptibility of the dot is plotted as a function of temperature in Fig.~\ref{Fig:suscept}.  Here
\begin{equation}   \label{eq:chi_0}
  \chi_0 =
  \frac{(2 g \mu_B)^{2}}{4 k_B T_K} .
\end{equation}
$\chi(T)$ increases from $0.033 \, \chi_0$ for $T = 20 \, T_K$ to $0.074 \, \chi_0$ for $T = 6 \, T_K$.

\section{Conclusions}
  \label{sec:conclusion}

We have demonstrated that the XX Heisenberg Hamiltonian of two one-dimensional spin 1/2 chains coupled to a quantum dot composed of two localized spin 1/2 magnets can be expressed as an Anderson model Hamiltonian for spin 1/2 fermions interacting with a spin 1/2 impurity. In this context, the chain index $\alpha$ plays a role of the fermion spin, where $\alpha = u, d$ stands for `spin-up' and `spin-down' states, respectively. At low temperatures, the impurity pseuodospin is screened by the chain pseudo-spins via the Kondo effect. The Kondo temperature $T_K$ was calculated and the temperature dependence of the entropy, specific heat and magnetic susceptibility of the dot were derived for $T \gg T_K$.

We plan to extend this calculation to the case where $H_0$ is an anisotropic XXZ Heisenberg Hamiltonian in a future publication.

\section*{Acknowledgement}

This research is dedicated to the memory of {\bf Professor Sergey Gredeskul} who was murdered by Hamas terrorists in his home in Ofakim (Israel) on October 7 2023.
 \\

\underline{Short Obituary written by IK}:

\begin{center}
\begin{tabular}{l}
``There are no goodbyes for us. \\
Wherever you are, you will be \\
in our hearts.'' \\
{\it Mahatma Gandhi }
\end{tabular}
\end{center}

The loss of a friend and my teacher is a difficult experience, especially when it is unexpected and untimely. I was deeply saddened and dismayed to learn that Professor Sergey Gredeskul and his wife, Viktoria, were brutally murdered by Hamas operatives in their home in Ofakim.

I first met Professor Gredeskul in 2000, at the beginning of my doctoral studies. Professor Gredeskul (together with Professor Yshai Avishai and Professor Konstantin Kikoin) was my thesis advisor. I was a newcomer to Israel, and he helped me both with my scientific research and with the everyday challenges faced by newcomers.   I am grateful for the opportunity to communicate with him and learn from him. He was a kind person and an exemplary mentor. May his memory be blessed.


\end{document}